\documentclass[aps,showpacs,a4paper,floatfix,twocolumn,prc,amsmath,amssymb]{revtex4}
\usepackage{amsmath,lscape,epsfig}
\usepackage{amsfonts}
\usepackage{amsmath}
\usepackage{amssymb}
\usepackage{slashed}
\usepackage{color,soul}
\usepackage{graphicx}
\usepackage{epsfig}
\usepackage{epstopdf}
\usepackage{ulem}
\usepackage{morefloats}
\usepackage{rotating}
\usepackage{relsize}
\usepackage{color}      
\definecolor{mypink}{RGB}{219, 48, 222}
\definecolor{brown}{RGB}{200,150, 50}

\usepackage{url}

\usepackage{amsmath}

\begin{document}

\title{Stability of the neutron-proton-electron matter under strong magnetic 
fields: the covariant Vlasov approach}

\author{Sidney Avancini$^{1,2}$}
\author{B. P. Bertolino$^2$}
\author{Aziz Rabhi$^{1,3}$}
\author{Jianjun Fang$^{1,4}$}
\author{Helena Pais$^1$}
\author{Constan\c ca Provid{\^e}ncia$^1$}

\affiliation{$^1$CFisUC, Department of Physics, University of Coimbra, 3004-516 Coimbra, Portugal.\\
$^2$Departamento de F\'{\i}sica, Universidade Federal de Santa Catarina, Florian\'opolis, 
SC, CP. 476, CEP 88.040-900, Brazil. \\
$^3$ University of Carthage, Avenue de la R\'epublique BP 77 -1054 Amilcar, Tunisia. \\
$^4$ School of Physics and Physical Engineering, Qufu Normal University, 273165 Qufu, China.
}

\begin{abstract}
The neutron-proton-electron({\it npe}) matter under a strong magnetic field is studied in the context of
the covariant Vlasov approach. We use a Walecka-type hadronic model and the dispersion 
relations for the longitudinal and transverse modes are obtained. The instability regions for longitudinal 
and transverse modes are also studied. The crust-core transition of a
magnetized neutron star is discussed.
\end{abstract}

\maketitle

\section{Introduction}

Neutron stars are astrophysical objects with extreme properties,
and very high densities in their interior, which could be as high
as one order of magnitude larger than nuclear matter at saturation
densities,  very intense surface  magnetic fields, that can reach
10$^{15}$ G in the case of magnetars \cite{duncan}, and very large isospin asymmetries.
These properties transform neutron stars into effective laboratories
for understanding the strong force. Objects as soft $\gamma$-ray
repeaters and anomalous X-ray pulsars have been identified as
magnetars carrying very strong  surface magnetic fields
\cite{sgr}. They have a slow rotation period, $\sim 1-12$ s, and this has been interpreted
as a signature of the existence of an amorphous and heterogeneous
layer, with a  high electrical resistivity, close to the crust-core
transition which causes the decay of the magnetic field \cite{pons13}.

The inner crust of neutron stars is characterized by subsaturation
nuclear densities and it  is formed by a lattice of
neutron rich clusters imbedded in a neutron and electron gas
background. In fact, at subsaturation densities, nuclear matter is
characterized by a liquid-gas phase transition. As a consequence of  long-range Coulomb repulsion and
short-range nuclear attraction this transition originates the formation
of clusterized matter that, depending on the density, may have exotic
geometries, and which are known as nuclear pasta \cite{pasta}.  At the
bottom of the inner crust, where these clusters occur, the neutron
star core sets in. Understanding how the magnetic field may affect the
localization of the core onset has been the objective of recent
works \cite{Fang16,Fang16a,Fang17}, where the crust-core transition in
the presence of strong magnetic fields was studied within the
calculation of the dynamical spinodal section \cite{umodes06}. From the
linearization of the Vlasov equation for the distribution function of
protons, neutrons and electrons, the dispersion relation for the
propagation of longitudinal models along the magnetic field was
calculated. The spinodal section was identified with the surface in
phase space where the density mode frequencies are zero. The main
conclusions of those works were that the magnetic field shifts the
onset of the outer core to larger densities, the crust-core transition
is a region of finite density thickness, the density location of this
transition region formed by alternating stable and unstable layers was
quite sensitive to the slope of the symmetry energy, and temperatures of the order of $\sim0.5-1$ MeV might wash out
most of these effects.

One of the main objectives of the present work is to obtain the dispersion relations for
{\it npe} matter subjected to a strong external magnetic field
within the formalism of the  covariant Wigner function,  that allows us to calculate the
propagation of density modes in an arbitrary direction with respect to
direction of the magnetic field. The dispersion relations will be
applied to the calculation of the transverse and longitudinal collective modes, both stable and unstable.
The unstable 
modes determine the dynamical spinodal zones, which set boundaries for the inhomogeneous 
region of the magnetar crust.  This formalism is of particular
interest because it may be easily generalized
in order to calculate the electrical and thermal conductivity in the magnetized matter,
which is of fundamental importance for the study of the cooling of magnetars.
%
%
%
%
%
%
%
%
%
%
%
%
\section{Formalism}

The Lagrangian density using natural units, i.e., taking $c=\hbar=$1, can be written as
\begin{equation}
{\cal L}=\sum_{j=p,n,e} {\cal L}_j + \cal L_\sigma + {\cal L}_\omega +
{\cal L}_\rho + {\cal L}_{\omega \rho } + {\cal L}_{A} \ ,
\label{lag}
\end{equation}
with
\begin{equation}
{\cal L}_j=\bar \psi^{(j)}\left[\gamma^\mu i D_\mu^{(j)}-M^*_j \right]\psi^{(j)} ,\nonumber
\end{equation}
where the covariant derivative 
is defined as, $ iD^{(j)}_{\mu}~= ~ i \partial_{\mu} - 
{\cal V}^{(j)}_{\mu} $,
where $j=(n,p,e)$, stands for the neutron, proton and electron,
\begin{equation}
 {\cal V}^{(j)}_{\mu} = 
 \left\{
       \begin{array}{l}
         g_v V_\mu  + \frac{g_\rho}{2}\, \vec{b}_\mu+ e\, A_\mu \ ~ ,~j=p \\  
         g_v {V}_\mu -\frac{g_\rho}{2}\, \vec{b}_\mu \ ~ ,~j=n \\
         - e\, A_\mu \ ~ ,~ j=e
     \end{array}    \right. \ ,  
\end{equation}
$M^*_p=M^*_n=M^*=M-g_s\phi(x),$ $M^*_e=m_e$ and
$e=\sqrt{4\pi/137}$ is  the electromagnetic coupling constant.
For the nuclear matter parameters, we will consider the NL3 parametrization, 
 \cite{nl3}, and the FSU parametrization \cite{fsu}.
The meson and photon contributions in eq. (\ref{lag}) are given by
\begin{eqnarray}
\mathcal{L}_{{\sigma }} &=&\frac{1}{2}\left( \partial _{\mu }\phi \partial %
^{\mu }\phi -m_{s}^{2}\phi ^{2}-\frac{1}{3}\kappa \phi ^{3}-\frac{1}{12}%
\lambda \phi ^{4}\right) \ , \nonumber \\
\mathcal{L}_{{\omega }} &=&\frac{1}{2} \left(-\frac{1}{2} \Omega _{\mu \nu }
\Omega ^{\mu \nu }+ m_{v}^{2}V_{\mu }V^{\mu }
+\frac{1}{12}\xi g_{v}^{4}(V_{\mu}V^{\mu })^{2} 
\right) \ , \nonumber \\
\mathcal{L}_{{\rho }} &=&\frac{1}{2} \left(-\frac{1}{2}
{\vec{B}}_{\mu \nu }\cdot {\vec{B}}^{\mu
\nu }+ m_{\rho }^{2} \vec{b}_{\mu }\cdot \vec{b}^{\mu } \right)   \ , \nonumber \\
\mathcal{L}_{\omega \rho } &=& \Lambda_v g_v^2 g_\rho^2 V_{\mu }V^{\mu }
\vec{b}_{\nu }\cdot \vec{b}^{\nu }   \nonumber \\
\mathcal{L}_{A} &=&-\frac{1}{4} F_{\mu \nu }F ^{\mu \nu }~, \label{mesonlag}
\end{eqnarray}
where $\Omega _{\mu \nu }=\partial _{\mu }V_{\nu }-\partial _{\nu }V_{\mu }$, 
$\vec{B}_{\mu \nu }=\partial _{\mu }\vec{b}_{\nu }-\partial _{\nu }
\vec{b}
_{\mu }-\Gamma_{\rho }(\vec{b}_{\mu }\times \vec{b}_{\nu })$ and 
$F_{\mu \nu }=\partial _{\mu }A_{\nu }-\partial _{\nu }A_{\mu }$.
The parameters $\kappa$, $\lambda$ and $\xi$ are
self-interacting couplings and the $\omega-\rho$ coupling
$\Lambda_v$ is included to soften the density dependence of the
symmetry energy above saturation density.
From the Euler-Lagrange equations, one obtains the Dirac equation for the fermion fields:
\begin{eqnarray}
 i \gamma^\mu D^{(j)}_{\mu}~\psi^{(j)} = M^{\star}_{j} ~\psi^{(j)} \ , \label{dirac1}
\end{eqnarray}
and its conjugate equation:
\begin{eqnarray}
 \bar{\psi}^{(j)} i D^{\dagger (j)}_{\mu}\gamma^\mu ~= - M^{\star}_{j} ~\bar{\psi}^{(j)} \, 
 \label{dirac2}
\end{eqnarray}
where $ iD^{\dagger (j)}_{\mu} = ~ i\overleftarrow{\partial}_{\mu} + {\cal V}^{(j)}_{\mu} $ .
In this section, we will discuss how the Vlasov equation for a hadronic system in a 
strong magnetized medium 
is obtained from general transport equations.
Our formalism is based on the covariant Wigner function under strong magnetic fields. 
The transport equations 
using Wigner functions have been developed in the context of quantum electrodynamics \cite{Vasak,Heinz}, 
relativistic heavy ions \cite{Mosel1,Mosel2} and quantum chromodynamics \cite{Elze}. 
The application of the Wigner 
function for 
a relativistic electron gas in a strong magnetic field was done in \cite{Hakim1,Hakim2,Hakim3,Hakim4}.
We will present in this section only the main results related to the transport theory in order 
to keep this paper minimally self-contained.
We will focus on the new technical details that appear when using the covariant Wigner function 
for the description of {\it npe} matter subjected to strong magnetic fields
since, in the present context and to our knowledge, this has not yet 
been done. It will be shown that both the longitudinal and transverse modes, 
which consist in small oscillations along and perpendicular to the external magnetic field, 
can be obtained in a systematic way by generalizing magnetized plasma physics techniques. 
The present formalism is adequate for the study of collective modes and, besides, 
opens the possibility to study thermal and electrical conductivity in {\it npe} matter.
\subsection{Covariant Wigner function}
We define the Wigner covariant matrix operator \cite{Heinz} as:
\begin{equation}
 \hat{W}^{(j)}_4(x,p) = \int d^4y ~e^{-i p \cdot y} ~ \Phi^{(j)}_4 (x,y) \ , \label{wigner}
\end{equation}
where $x$ and $y$ are 4-vectors, $j=(n,p,e)$, and 
\begin{eqnarray}
 && \Phi^{(j)}_4 (x,y)_{\alpha \beta} \equiv 
 \bar{\psi}^{(j)}_\beta(x) e^{ \frac{y}{2} \cdot D^{\dagger (j)} } 
   e^{ - \frac{y}{2} \cdot D^{(j)} } \psi^{(j)}_\alpha (x)     \nonumber \\  
  && = \bar{\psi}^{(j)}_\beta(x_+) e^{-i y \cdot \int_{-1/2}^{1/2}~ds~ {\cal V}^{(j)}(x+ys)  }
  ~\psi^{(j)}_\alpha (x_-) \ .   \label{phi0}
\end{eqnarray}  
In the latter equation, $x_\pm=x \pm \frac{1}{2}y$, and
the phase factor contains a line integral which is to be
calculated along a straight line path from 
$x_-$  to $x_+$, since in this case the gauge invariance of 
the Wigner function is guaranteed \cite{Vasak}. If one defines the canonical, $\hat{p}_\mu$,
and kinetic, $\hat{\Pi}_\mu$, momentum operators as:
\begin{equation}
  \hat{p}_\mu=\frac{1}{2i}(\overleftarrow{\partial}_{\mu} -\partial_\mu ) ~~, ~~
  \hat{\Pi}_\mu = \frac{1}{2i}( D^{\dagger (j)}_\mu - D^{(j)}_\mu ) = \hat{p}_\mu - 
  {\cal V}^{(j)}_\mu \ ,
   \nonumber
\end{equation}
then, from eqs. (\ref{wigner},\ref{phi0}), the Wigner matrix can be rewritten as:
\begin{equation}
 \hat{W}^{(j)}_4(x,p)_{\alpha\beta} = (2\pi)^4 \bar{\psi}^{(j)}_\beta(x) 
 \delta^4 (\hat{\Pi} (x) - p) 
 \psi^{(j)}_\alpha (x)
 \ . \nonumber
\end{equation}
From the Dirac equations, (\ref{dirac1},\ref{dirac2}), in a rather 
cumbersome calculation \cite{Vasak,Heinz}, it is possible to derive the following equation for
$\hat{W}_4(x,p)$:
\begin{equation}
 \left[ \gamma^\mu (\hat{\Pi}_\mu + \frac{i\hbar}{2} {\cal D}_\mu) - 
 M^\star (x - \frac{i\hbar}{2} \partial_p)   \right]\hat{W}_4(x,p)=0 \ , \label{wignereq1}
\end{equation}
where
\begin{eqnarray}
 {\cal D}_\mu=\partial_{x \mu} - \int_{-1/2}^{1/2}~ds~ {\cal F}_{\mu \nu} (x+ i \hbar s \partial_p) 
 \partial_p^\nu
 \ ,
\end{eqnarray}
\begin{eqnarray}
 \hat{\Pi}_\mu= p_{\mu} + i\hbar \int_{-1/2}^{1/2}~ds~ s 
   {\cal F}_{\mu \nu} (x+ i \hbar s \partial_p) \partial_p^\nu ~,
\end{eqnarray}
and     $~{\cal F}_{\mu \nu}^{(j)}= \partial_{x \mu} {\cal V}_\nu^{(j)} 
           - \partial_{x \nu} {\cal V}_\mu^{(j)}$  . { In
           Eq. \ref{wignereq1}
$$M^\star (x - \frac{i\hbar}{2} \partial_p) =\sum_{n=0}^\infty
\frac{1}{n!}\left(\frac{-i\hbar}{2}\right)^n \Delta^n M^\star (x),$$
with  $ \Delta=\partial_x\cdot\partial_p,$
and $\Delta M^\star
(x)=\left(\frac{\partial}{\partial x_\mu} M^\star (x)\right) \frac{\partial}{\partial p_\mu}.$
Note that
             $\partial_{x \mu}\equiv \partial_{\mu}$. The first
             designation will be used if necessary to distinguish from
           the 4-gradient with respect to $p$. }

In the latter equations, $\hbar$ is explicitly shown, since our aim is to consider 
approximations of $O(\hbar)$ in order to derive the semiclassical Vlasov equation.
From here on, we will omit the particle index, $(j)$, whenever this does not cause
confusion. Notice that eq. (\ref{wignereq1}), when $\hbar \rightarrow 0$, can be written as:
\begin{equation}
 (p^2 - M^\star(x)^2)\hat{W}_4(x,p)=0 ~ \, \nonumber
\end{equation}
or 
\begin{equation}
({p^0}^2 - 
              ({\vec p}^{~2}+M^\star(x)^2) )~ \hat{W}_4(x,p)=0  \nonumber \ .
\end{equation}
This result means that the Wigner operator is on the mass shell, thus, the component
of $O(\hbar^0)$ of $\hat{W}_4$ can be written as:
\begin{equation}
 \hat{W}_4(x,p)=\hat{W}^{(+)(0)}_4 \delta (p_0 - E_p) +
 \hat{W}^{(-)(0)}_4 \delta (p_0 + E_p) \ ,  \label{onmass}
\end{equation}
where $E_p=\sqrt{{\vec p}^{~2}+ M^\star(x)^2}$.
Here, we will discuss briefly how the Vlasov equation may be obtained from the Wigner operator.
First \cite{Vasak}, we expand the Wigner $4\times 4$ matrix operator in terms of the Clifford algebra:
\begin{eqnarray}
 \hat{W}_4(x,p)&=&\frac{1}{4} \left[ F(x,p) I + i\gamma^5 P(x,p) +\gamma^\mu F_\mu(x,p) \right.
                                             \nonumber \\
&& \left. + \gamma^\mu \gamma^5 \Omega_\mu (x,p) + \frac{1}{2} \sigma^{\mu \nu} 
                           S_{\mu \nu}(x,p)
  \right]  \ ,  \label{cliff}
\end{eqnarray}
where $\sigma^{\mu \nu}= \frac{1}{2}( \gamma^\mu \gamma^\nu-\gamma^\nu \gamma^\mu)$ .
After substituting the Clifford expansion of $\hat{W}_4$ in eq. (\ref{wignereq1}), we 
calculate the traces in Dirac space by multiplying the resulting expression by each 
one of the sixteen Clifford algebra elements. Thus, we obtain sixteen 
complex equations \cite{Heinz}. Next, we consider the energy average of the Wigner function, since 
the Vlasov equation is usually written in terms of the Cartesian momentum vector, $\vec{p}$,
instead of the 4-vector $p^\mu$. For convenience, we define $\hat{W}_3(x,\vec{p})$ as:
\begin{equation}
 \hat{W}_3(x,\vec{p})=\int d p_0 \hat{W}_4 (x,p^0,\vec{p})~\gamma^0 \ .
\end{equation}
Of course, one obtains again sixteen equations relating the components of the $\hat{W}_3$ just defined.
It is possible to show \cite{Heinz}, after a long and tedious calculation, that at $O(\hbar)$ 
one obtains the following generalized Vlasov equation:
 \begin{eqnarray}
  && \partial_t f_{\pm} \pm \vec{v} \cdot \nabla_x f_{\pm} +
  (\vec{{\cal E}} \pm \vec{v} \times \vec{\cal B})\cdot \nabla_p f_{\pm} \nonumber \\
  && \mp \frac{M^\star}{E_p} \nabla_x M^\star \cdot \nabla_p f_{\pm} = 0  \ ,
  \label{vlasov}
 \end{eqnarray}
where $\vec{v}=\vec{p}/E_p$ and
\begin{eqnarray}
 && {\cal E}_i^{(j)} = {\cal F}_{0i}^{(j)} \mp \frac{M_j^\star(x)}{E_p^{(j)}} 
 \nabla_{x,i}~ M_j^\star(x) ~\ , \  j=p,n \nonumber \ , \\
 && {\cal E}_i^{(e)} = {\cal F}_{0i}^{(e)}  \nonumber \ , \\
 && {\cal B}^{(j)}_i ~=~\epsilon_{ilm} \partial_{x,l} {\cal V}^{(j)}_m ~,~j=p,n,e~\ ,
\end{eqnarray}
$i,l,m=1,2,3$ and 
\begin{equation}
 f_{\pm}(x,\vec{p}) = \int dp_0 F_{0\pm}(x,p) =\int dp_0 Tr[ \gamma^0 
 \hat{W}^{(\pm)(0)}_4] \ . \nonumber \\
\end{equation}
An important remark is in order: the Vlasov equation can be derived in other approaches, 
for instance, in Refs. \cite{nielsen89,nielsen91}, this equation has been obtained 
using a Hamiltonian formalism. The main advantage of the Wigner approach is that one has a 
systematic method for obtaining the particle equilibrium distribution functions, 
which is of fundamental
importance for the description of systems in a magnetized medium.
\subsection{Wigner equilibrium function in a magnetized medium}
The equilibrium Wigner distribution function is defined as:
\begin{equation}
 f_{\pm}^{(0)(j)}(x,\vec{p}) =\frac{1}{(2\pi)^4} \int dp_0 Tr[ \hat{\rho}~ \gamma^0 
 \hat{W}^{(\pm)(0)}_4] \ , \nonumber \\
\end{equation}
where the density matrix operator is given by:
\begin{equation}
 \hat{\rho}= \frac{1}{Z} e^{-\beta (\hat{H}-\sum_j \mu_j \hat{N}_j )} \ ,
\end{equation}
where $Z$ is the partition function, $\beta=1/T$, and the trace is to be evaluated 
simultaneously in the Dirac space and in any convenient set of basis states, for example,
the Johnson-Lippmann \cite{john}. Of course, $f^{(0)}_{\pm}(x,\vec{p})$ satisfies the Vlasov 
equation by its construction.
Substituting the explicit form of the Wigner operator, eq. (\ref{wigner}), and the on-mass-shell 
constraint, eq. (\ref{onmass}), one obtains:
\begin{eqnarray}
 &&f^{(0)(j)}_{\pm}(x,\vec{p}) =\int \frac{d p_0}{(2\pi)^4} Tr[ \hat{\rho}~ \gamma^0
                             \int d^4 y      e^{-i p\cdot y}\nonumber \\ 
 &&\bar{\psi}^{(j)}_{\pm}(x_+) e^{-i y \cdot \int_{-1/2}^{1/2}~ds~ {\cal V}^{(j)}(x+ys)  }
  ~\psi^{(j)}_{\pm} (x_-)    ] \ ,  \label{wignereq}
\end{eqnarray}
where $\psi^{(j)}_{\pm}(x)$, $j=(n,p,e)$, are the positive and negative components of the
fermionic Dirac fields associated to the neutron, proton and electron. 
From here on, we restrict our calculations to the mean field approximation. Therefore,   
only the time component of the $\omega$ and $\rho$ vector meson fields are 
different from zero at equilibrium:
\begin{equation}
V^{(0)}_\mu=V_0 \delta_{\mu 0} ~,~ b^{(0)}_\mu=b_0 \delta_{\mu 0}  \ . 
\end{equation}
We choose the Landau gauge for the vector potential:
 $A^{(0)}_\mu= B x_2 \delta_{\mu 3}$, thus the external strong magnetic 
field is $\vec{B}=B\hat{e}_3$ and $\nabla \cdot \vec{B}$ = 0.
The integral in the phase factor, eq.(\ref{wignereq}), can 
be easily evaluated resulting in ${\cal V}^{(0)(j)}_{\mu}$, thus:
\begin{eqnarray}
 &&f_{\pm}^{(0)(j)}(x,\vec{p}) =\int \frac{d p_0}{(2\pi)^4} \int d^4 y  
            Tr[ \hat{\rho}~ \gamma^0
                                 e^{-i (p+{\cal V}^{(0)(j)}) \cdot y}\nonumber \\ 
 &&\bar{\psi}^{(j)}_{\pm}(x_+) 
  ~\psi^{(j)}_{\pm} (x_-)    ] \ . \nonumber \\ \label{wignereq3}
\end{eqnarray}
For the calculation of the equilibrium function, one has to insert the Dirac field
operator in the Fock space in the last expression and to perform the integrals indicated 
in eq. (\ref{wignereq3}). A similar calculation has been done in \cite{Hakim1} 
for a magnetized electron gas and can be promptly generalized for the present situation resulting
in the following expression:
\begin{eqnarray}
 &&f_{\pm}^{(0)(j)}(\vec{p}) =  \sum_{n=0}^\infty  
            \left[ L_{n}(2w^2)-L_{n-1}(2w^2) \right] \nonumber \\ 
 &&\times \frac{ (-1)^{n} ~e^{-w^2}}{1+\exp\left[ \beta(E_n \mp \tilde{\mu}^{(j)})\right] }
 \ , j=p,e \nonumber \\ 
 \label{wignereq4}
\end{eqnarray}
where $w^2=\frac{p_\perp^2}{eB}=\frac{p_1^2+p_2^2}{eB}$, $L_n(x)$ are the Laguerre 
polynomials, 
$\tilde{\mu}^{(e)} = \mu_e $, and $\tilde{\mu}^{(p)} = \mu_p - {\cal V}^{(0)(p)}_0 $ 
are the effective chemical potential of the electron and proton respectively, 
$ E_n=\sqrt{p_3^2+{M^{\star}_j }^2 + 2eBn }$ { where $n$ is the
  Landau level label,}
and we have defined $L_{-1}(x)$=0. Since in this work we are only interested 
in systems at zero temperature, $T=0$, one can rewrite the last equation as:
\begin{eqnarray}
 &&f^{(0)(j)}(\vec{p}) = \sum_{n=0}^\infty  
            \left[ L_{n}(2w^2)-L_{n-1}(2w^2) \right]  \nonumber \\ 
 &&\times (-1)^{n}~ e^{-w^2} \theta( \tilde{\mu}^{(j)} -E_n)
 \ ,   \label{wignereq5}
\end{eqnarray}
where the Heaviside function $\theta(x)$ was used. The limit $T \to 0$
rules out the negative energy part of the distribution function, hence we will 
omit from here on the redundant plus signal in the symbol  $f_{+}^{(0)(j)}$.
The electron or proton equilibrium densities can be readily calculated from 
their corresponding distribution functions. For example, the electron density
reads from eq. (\ref{wignereq5}):
\begin{eqnarray}
 &&\rho^{(0)}_e=\frac{2}{(2\pi)^3} \int d^3 p~ f^{(0)(e)}(\vec{p}) \nonumber \\
 &&=\frac{2}{(2\pi)^3} \int_{-\infty}^{\infty}~d p_\parallel \int_{0}^{\infty} ~d p_\perp p_\perp
 \int_0^{2\pi} d \Phi ~f^{(0)(e)}(\vec{x},\vec{p},t) \nonumber \\
&& =\frac{eB}{\pi^2} \sum_{n=0}^\infty  
\int_{0}^{\infty}d p_\parallel \int_{0}^{\infty} d w ~w 
            \left[ L_{n}(2w^2)-L_{n-1}(2w^2) \right]\nonumber \\ 
 &&\times  e^{-w^2} (-1)^{n}  \theta(\mu_e - E_n) \ .
\end{eqnarray}
The last expression can be easily calculated in cylindrical coordinates, 
since the integral in $w$ can be found in tables of integrals \cite{grad},
and the integral in $p_\parallel$=$p_3$ is calculated using the Heaviside
function. Therefore:
\begin{equation}
 \rho^{(0)}_e =\frac{eB}{2\pi^2} \sum_{n=0}^{n_{max}} g_n~ p_F^{(e)}(n) \label{densie} \ ,
\end{equation}
with $n_{max}=[\frac{\mu_e^2-m_e^2}{2eB}]$ where [...] 
stands for the floor function, which
gives the greatest integer that is less than or equal to x. $g_n$ is a degeneracy 
factor which is 1 for $n=0$ and 2 for $n \geq 0$, and $p_F^{(e)}(n)=\sqrt{\mu^2_e - m_e^2-2eBn}$.
The density of electrons just obtained coincides with the usual one \cite{broderick}.
The neutron distribution function is also the standard one: 
$f^{(0)(n)}(\vec{p}) = \theta(p_F^{(n)~2}-\vec{p}^{~2})$. Notice that the 
normalization factor of the distribution functions are determined according to the definition 
of the currents as given in the next section.
\subsection{Dispersion relations}

The 4-current is given by:
\begin{equation}
 J^{(j)}_\mu(x)=\frac{2}{(2\pi)^3} \int \frac{d^3 p}{p^0} p_\mu 
 (f^{(j)}_+(x,\vec{p}) - f^{(j)}_-(x,\vec{p}) ) \label{curr} \ ,
\end{equation}
where $j=(n,p,e)$ and {$p^0=E_j$=$\sqrt{\vec{p}^{~2}+M_j^{\star~2}}$}. It follows 
from the Vlasov equation, eq. (\ref{vlasov}),
the following 4-current conservation law: $\partial^\mu  J^{(j)}_\mu(x)$ = 0.

The dispersion relations will be obtained considering a small perturbation from
equilibrium of the distribution functions, where the equilibrium 4-current is given by:
\begin{equation}
  J^{(0)(j)}_\mu = \frac{2}{(2\pi)^3} \int \frac{d^3 p}{p^0} p_\mu 
 (f^{(0)(j)}_{+}(\vec{p}) -f^{(0)(j)}_{-}(\vec{p}))   \ . \nonumber
\end{equation}
Next, we consider only the zero-temperature case ,$T=0$. Hence, we can write:
\begin{equation}
  f^{(j)}(\vec{x},\vec{p},t) = f^{(0)(j)}(\vec{p}) + \delta f^{(j)} \,
  \label{delphi}.
\end{equation}
The small perturbation of the distribution functions, $f^{(0)(j)}$, around their equilibrium 
values, will generate perturbations on the fields:
\begin{eqnarray}
&& \phi = \phi^{(0)}+\delta \phi ~,~V_\mu=V_\mu^{(0)}+\delta V_\mu ~,~
 b_\mu=b_\mu^{(0)}+\delta b_\mu~, \nonumber \\
 &&~A_\mu=A_\mu^{(0)}+ \delta A_{\mu}~ , \label{smalldev}
\end{eqnarray}
and cause a corresponding perturbation of the equilibrium 4-current,
\begin{equation}
  J^{(j)}_\mu(x) = J^{(0)(j)}_\mu
   + \delta J^{(j)}_\mu \ ,
\end{equation}
with
\begin{equation}
  \delta J^{(j)}_\mu = \frac{2}{(2\pi)^3} \int \frac{d^3 p}{E^{(0)}_j}
    ~p_\mu ~\delta f^{(j)}   \ , \label{curre}
\end{equation}
where we have used the notation { $E^{(0)}_j=\sqrt{\vec{p}^{~2}+{M_j^{\star (0)}}^2}$} , $j=(n,p,e)$,
with $M_p^{\star (0)}$ = $M_n^{\star (0)}$ = $M-g_s \phi^{(0)}$ and $M_e^{\star (0)}$=$m_e$. 
The scalar density is given by the expression:
 \begin{equation}
  \rho_s^{(j)}= \frac{2}{(2\pi)^3} \int \frac{d^3 p}{E_j} {M_j^\star}
 f^{(j)}(x,\vec{p})    \ .
 \end{equation}
The small perturbation of the proton and neutron scalar densities have to be calculated with 
care \cite{avancini05}, since ${M_j^\star}= M-g_s \phi(x)$ is position-dependent, 
resulting in the following expression:
\begin{equation}
 \rho_s^{(j)}= \rho_s^{(0)(j)}+ \delta \rho_s \ ,
\end{equation}
with
\begin{equation}
  \rho_s^{(0)(j)}= \frac{2}{(2\pi)^3} \int \frac{d^3 p}{E^{(0)}_j} {M_j^\star}^{(0)} 
 f^{(0)(j)}(\vec{p})    \ , 
\end{equation}
and $ \delta \rho_s^{(j)}= \delta\tilde{\rho}_s^{(j)}+ g_s ~d\rho_s^{(0)(j)}~ \delta \phi$ ,
with:
 \begin{equation}
  \delta \tilde{\rho}_s^{(j)}= \frac{2}{(2\pi)^3} \int \frac{d^3 p}{E^{(0)}_j} 
  {M_j^\star}^{(0)} \delta f^{(j)}    \ ,  \label{densca}
 \end{equation}
 and
 \begin{equation}
  d\rho_s^{(0)(j)} = - \frac{2}{(2\pi)^3} \int d^3 p 
  \frac{{\vec{p}}^2} {{E^{(0)}_j}^3}   f^{(0)(j)}    ~.
 \end{equation}
\noindent
After substituting eq. (\ref{delphi}) in the Vlasov equation, eq. (\ref{vlasov})
retaining only terms of the first order in $\delta f^{(j)}$, one obtains:
 \begin{eqnarray}
  && \partial_t \delta f^{(j)} + \vec{v} \cdot \nabla_x \delta f^{(j)} +
  \vec{v} \times ( \nabla_x \times \vec{\cal V}^{(0)(j)} ) \cdot \nabla_p \delta f^{(j)} \nonumber \\
  && + \left[ \vec{v} \times \nabla_x \times (\vec{\cal V}^{(0)(j)}  + 
\delta \vec{\cal V}^{(j)} ) + g_s \frac{{M_j^\star}^{(0)}}{{E^{(0)}_j}} \nabla_x \delta \phi 
\right. \nonumber \\
 && \left.   - \partial_t ~\delta \vec{{\cal V}}^{(j)} -
 \nabla_x \delta {\cal V}_0^{(j)}  \right]  \cdot 
 \nabla_p f^{(0)(j)}    
   = 0 \ ,
  \label{dvlasov}
 \end{eqnarray}
where $\vec{v}=\vec{p}/E^{(0)}_j$ , $j=p,e$ ( for the electrons $g_s=0$). 
The last equation can be further simplified noting that the equilibrium distribution 
function, eq. (\ref{wignereq5}), is a function of $p_\parallel$ and $p_\perp$ only, hence,
it is useful to employ cylindrical coordinates $(p_\perp,\Phi,p_\parallel)$ to 
rewrite eq. (\ref{dvlasov}) as:
 \begin{eqnarray}
  && \partial_t \delta f^{(j)} + \vec{v} \cdot \nabla_x \delta f^{(j)} - \frac{Q_j B}{E^{(0)}_j}
  \frac{\partial}{\partial \Phi}  \delta f^{(j)} \nonumber \\
  && + \left[ \vec{v}  \times (\nabla_x \times \delta \vec{\cal V}^{(j)} ) + 
  g_s \frac{{M_j^\star}^{(0)}}{{E^{(0)}_j}} \nabla_x \delta \phi  \right. \nonumber \\
 && \left.   - \partial_t ~\delta \vec{{\cal V}}^{(j)} -\nabla_x \delta {\cal V}_0^{(j)}  
 \right]  \cdot 
 \nabla_p f^{(0)(j)}    
   = 0 \ .
  \label{dvlasov2}
 \end{eqnarray}
 In the latter equation, $Q_j$ is the electric charge of the $j$ particle. Note also that this 
 equation is also valid for the neutron, if one takes $Q_n=0$.
Next, we obtain the dispersion relations, starting from the Fourier transform
of the small deviation from equilibrium of the fields and of the distribution functions:
 \begin{equation}
 \left\{
       \begin{array}{c}
          \delta f^{(j)} (\vec{x},\vec{p},t) \\
         \delta \phi (\vec{x},t) \\
         \delta {\cal V}^{(j)}_{\mu} (\vec{x},t)
       \end{array}    
 \right\}  = 
 \int d^3q ~d\omega 
  \left\{
       \begin{array}{c}
          \delta f^{(j)} (\vec{q},\omega,\vec{p}) \\
          \delta \phi (\vec{q},\omega) \\
          \delta {\cal V}^{(j)}_{\mu}(\vec{q},\omega) 
       \end{array}    
 \right\}  e^{i(\omega t - \vec{q} \cdot \vec{x})} \ ,
 \end{equation}
and after substituting the last equation in the Vlasov equation, eq. (\ref{dvlasov2}), one 
obtains for $ \delta f^{(j)}(\vec{q},\omega,\vec{p})$ :
 \begin{eqnarray}
  && i (\omega - \vec{v} \cdot \vec{q} ) ~\delta f^{(j)} - \frac{Q_j B}{E^{(0)}_j}
  \frac{\partial}{\partial \Phi}  \delta f^{(j)} \nonumber \\
  && = i \left[ (\omega - \vec{v} \cdot \vec{q} ) ~\delta \vec{\cal V}^{(j)}  -
  \left( \delta {\cal V}_0^{(j)} - \vec{v} \cdot \delta \vec{{\cal V}}^{(j)}
   \right.   \right. \nonumber \\
 &&  \left. - g_s \frac{{M_j^\star}^{(0)}}{{E^{(0)}_j}} \delta \phi \right) \vec{q}\left.   
 \right]  \cdot 
 \nabla_p f^{(0)(j)}    
   = 0 \ .
  \label{dvlasov3}
 \end{eqnarray}
Here, we note that the last equation has an explicit dependence on the angle $\Phi$. In order to get rid
of this angular dependence, we will use some techniques which are well known in magnetized plasma physics, 
\cite{Kelly,Oberman}. We will adopt the same frame of reference used in \cite{Kelly} where:
\begin{eqnarray}
 && \vec{B}=B \vec{e}_3 ~,~\hat{e}_3 \equiv e_\parallel  \nonumber \\
 && \vec{q}= q_\perp \vec{e}_1 + q_\parallel \vec{e}_3 = (q_\perp,0,q_\parallel) \nonumber \\
 && \vec{p}= p_\perp \vec{e}_\perp + p_\parallel \vec{e}_3 = 
 (p_\perp \cos \Phi,p_\parallel \sin \Phi, p_\parallel )  ~ .
\end{eqnarray}
The angular dependence of $\delta f^{(j)} (\vec{q},\omega,\vec{p})$ = 
$\delta f^{(j)} (\vec{q},\omega,p_\perp,\Phi,p_\parallel)$ can be separated by using the 
Oberman-Ron \cite{Oberman} transform:
\begin{eqnarray}
 &&\delta f^{(j)} (\vec{q},\omega,\vec{p}) = \nonumber \\
 && e^{i b \sin \Phi} \sum_{m=-\infty}^{\infty} e^{-im\Phi} J_m(b) 
 \delta f^{(j)}_m (\vec{q},\omega,p_\parallel,p_\perp) \ , \label{trans} 
\end{eqnarray}
and its inverse transform:
\begin{eqnarray}
 &&\delta f^{(j)}_m (\vec{q},\omega,p_\parallel,p_\perp) = \nonumber \\
 &&\frac{1}{2\pi J_m(b)} \int_0^{2\pi} d\Phi e^{-i b \sin \Phi} e^{im\Phi} 
 \delta f^{(j)} (\vec{q},\omega,\vec{p}) ~ , \label{invtrans} 
\end{eqnarray}
where $J_m(b)$ is a Bessel function and $b$ is and arbitrary real constant. We substitute the 
Oberman-Ron transform, eq. (\ref{trans}), in the Vlasov equation, eq. (\ref{dvlasov3}),
with $b=-\frac{q_\perp p_\perp}{Q_j B}$, $j=(p,e)$, and integrate in $\Phi$ the resulting
expression multiplied by the factor $\exp[-i(b\sin \Phi-m\Phi)]$. After some straightforward,
but tedious calculation, we finally obtain, using eq. (\ref{invtrans}), the following expression:
\begin{eqnarray}
&&  \delta f^{(j)} (\vec{q},\omega,\vec{p})= e^{i b \sin \Phi} \sum_{m=-\infty}^{\infty}
  e^{-im\Phi}   \nonumber \\
&&\times \left\{  \frac{m}{b} J_m(b)\left[\omega D_\perp^{(j)}- \frac{q_\parallel}{E^{(0)}_j}
(p_\parallel D_\perp^{(j)} -p_\perp D_\parallel^{(j)}) \right]   \delta {\cal V}^{(j)}_x \right.  \nonumber \\ 
&&  + i J_m^\prime (b)\left[\omega D_\perp^{(j)}- \frac{q_\parallel}{E^{(0)}_j}
(p_\parallel D_\perp^{(j)} -p_\perp  D_\parallel^{(j)}) \right]
                                                                 \delta {\cal V}^{(j)}_y   \nonumber \\   
&&  + J_m (b)\left[\omega D_\parallel^{(j)}- \frac{q_\perp }{E^{(0)}_j}
(p_\perp D_\parallel^{(j)} -p_\parallel  D_\perp^{(j)}) \frac{m}{b} \right]
                                                                 \delta {\cal V}^{(j)}_z   \nonumber \\ 
&& \left. - J_m (b)\left[ q_\perp D_\perp^{(j)} \frac{m}{b} + q_\parallel D_\parallel^{(j)} \right] 
\left[ \delta {\cal V}^{(j)}_0 -\frac{g_s M_j^{\star(0)}}{E^{(0)}_j} \delta \phi \right] \right\}
\nonumber \\    
&& \times \left[ \omega - \frac{p_\parallel q_\parallel}{E^{(0)}_j} + \frac{Q_j B m}{E^{(0)}_j} 
\right]^{-1}
\label{dispvla}
\end{eqnarray}
where the prime of the Bessel function means its derivative with respect to $b$, and
\begin{equation}
 D_\perp^{(j)} = \frac{\partial}{\partial p_\perp} f^{(0)(j)}~,~ 
     D_\parallel^{(j)} = \frac{\partial}{\partial p_\parallel} f^{(0)(j)} ~ .
\end{equation}
The above expression is the key for the calculation of the dispersion relations in the 
presence of a strong external magnetic field. The Fourier transform of the small deviations from 
equilibrium of the current is given by:
\begin{equation}
   \delta J^{(j)}_\mu (\vec{x},t)=   \int d^3q ~d\omega
        ~  \delta J^{(j)}_\mu (\vec{q},\omega) 
        e^{i(\omega t - \vec{q} \cdot \vec{x})} \ .
\end{equation}
From eq.(\ref{curre}) it follows that:
\begin{equation}
   \delta J^{(j)}_\mu (\vec{q},\omega) = \frac{2}{(2\pi)^3} \int \frac{d^3 p}{E^{(0)}_j}
    ~p_\mu ~\delta f^{(j)}  (\vec{q},\omega,\vec{p})  \ , \label{delcurr}
\end{equation}
and the current conservation results in the relation:
\begin{equation}
 \partial^\mu J^{(j)}_\mu = 0 \Rightarrow \omega \delta J^{(j)}_0 + \vec{q} \cdot \delta 
\vec{J}^{(j)} =0 \ .
\end{equation}
The continuity equation also can be used to derive the important relation:
\begin{equation}
 \partial^\mu \delta {\cal V}^{(j)}_\mu = 0 \Rightarrow \omega \delta {\cal V}^{(j)}_0 + 
 \vec{q} \cdot \delta 
\vec{\cal V}^{(j)} =0 \ . \label{Vcons}
\end{equation}
\subsection{Dispersion relations for longitudinal and transverse density modes}
The dispersion relation for density perturbations follows from taking $\mu=0$ in eq. (\ref{delcurr}):
{\small{
\begin{eqnarray}
 && \delta \rho^{(j)} \equiv \delta J^{(j)}_0 (\vec{q},\omega)
 = \frac{2}{(2\pi)^3} \int d^3 p ~\delta f^{(j)}  (\vec{q},\omega,\vec{p})  \nonumber \\
 && = \frac{2}{(2\pi)^3} \int_0^\infty dp_\perp p_\perp \int_{-\infty}^\infty dp_\parallel 
 \int_0^{2\pi} d\Phi ~\delta f^{(j)}  (\vec{q},\omega,\vec{p}) \ ,
\end{eqnarray} }}
Then, substituting eq. (\ref{dispvla}) in the last expression and performing the
integration in $\Phi$,
one obtains:
{\small{
\begin{eqnarray}
&&  \delta \rho^{(j)} (\vec{q},\omega)=   \nonumber \\
&& S_{j} \left[ \frac{m}{b} J_m^2(b)\left(\omega D_\perp^{(j)}- \frac{q_\parallel}{E^{(0)}_j}
(p_\parallel D_\perp^{(j)} -p_\perp D_\parallel^{(j)}) \right)  \right] 
\delta {\cal V}^{(j)}_x   \nonumber \\ 
&&  + S_j \left[ i J_m(b)J_m^\prime (b)\left(\omega D_\perp^{(j)}- \frac{q_\parallel}{E^{(0)}_j}
(p_\parallel  D_\perp^{(j)} -p_\perp  D_\parallel^{(j)}) \right) \right]
                                                                 \delta {\cal V}^{(j)}_y   \nonumber \\   
&&  + S_j \left[J_m^2 (b)\left(\omega D_\parallel^{(j)}- \frac{q_\perp}{E^{(0)}_j}
(p_\perp  D_\parallel^{(j)} -p_\parallel  D_\perp^{(j)}) \frac{m}{b} \right) \right]
                                                                 \delta {\cal V}^{(j)}_z   \nonumber \\ 
&&  - S_j \left[ J_m^2 (b)\left( q_\perp D_\perp^{(j)} \frac{m}{b} + q_\parallel D_\parallel^{(j)} \right) 
\right] \delta {\cal V}^{(j)}_0  \nonumber \\
&&  + S_j \left[ \frac{J_m^2 (b)}{E^{(0)}_j} \left( q_\perp D_\perp^{(j)} \frac{m}{b} + 
q_\parallel D_\parallel^{(j)} \right) \right]  g_s {M_j^\star}^{(0)} ~ \delta \phi  \ ,
\label{disdensity}
\end{eqnarray} }}
where we have defined, as in Ref. \cite{Kelly}, the function:
{\small{
\begin{equation}
 S_j \left[ X \right] = \sum_{m=-\infty}^{\infty} \frac{1}{2\pi^2}\int_{-\infty}^{\infty}
 dp_{\parallel} \int_{0}^{\infty}
 \frac{dp_\perp~p_\perp   \left[  X \right]}
      {\omega -\frac{p_\parallel q_\parallel}{E^{(0)}_j} + \frac{Q_j Bm}{E^{(0)}_j} }~,
      \label{Lind}
\end{equation} }}
where $X$ is an arbitrary function of $p_\parallel$ and $p_\perp$. The next step in order to
obtain the dispersion relations is to substitute the mesonic deviations from equilibrium in 
the last equation by using the equations of motion. An additional equation involving the 
scalar density is necessary to determine the dispersion relation. From eq. (\ref{densca}),
it follows analogously to the particle density fluctuations that:
{\small{
\begin{equation}
\delta {\tilde \rho}^{(j)}_s 
 = \frac{2}{(2\pi)^3} \int d^3 p \frac{{M_j^\star}^{(0)} }{E^{(0)}_j}~
                                   \delta f^{(j)}  (\vec{q},\omega,\vec{p}) \ .
\end{equation} }}
Following the same reasoning used in the derivation of eq. (\ref{disdensity}), 
one obtains an analogous expression for the scalar density in terms of the meson fields and 
the generalized Lindhard function. In fact, the expression for $\delta {\tilde \rho}^{(j)}_s$ can 
be read from eq. (\ref{disdensity}) by including ${M_j^\star}^{(0)}/E^{(0)}_j $ in the 
integrands. In the appendix, we show, using the equations of motion for the meson and electromagnetic 
fields, that one can write the deviations from the equilibrium of these fields in terms
of the deviations of the corresponding densities. Therefore, substituting these fields
deviations in eq. (\ref{disdensity}), one obtains the dispersion relations.
\subsubsection{Longitudinal mode}
The longitudinal mode corresponds to small perturbations parallel to the magnetic field and 
the dispersion relations are obtained by taking
$q_\perp$=0, $q_\parallel=q$, $b=0$ and 
$\delta {\cal V}_x$ = $\delta {\cal V}_y$ =0 in eq.(\ref{disdensity}).
From the conservation law, eq. (\ref{Vcons}), $\partial^\mu {\cal V}^{(j)}_\mu=0$,
for the longitudinal modes we find:
\begin{equation}
 \omega \delta {\cal V}^{(j)}_0=\vec{q} \cdot \delta \vec{\cal V}^{(j)} = q \delta {\cal V}^{(j)}_z \ .
\end{equation}
The last result allows us to write $\delta {\cal  V}^{(j)}_z$ in terms of $\delta {\cal V}^{(j)}_0$, 
hence 
$\delta \rho^{(j)} (q,\omega)$ in eq. (\ref{disdensity}) reduces to:
{\small{
\begin{eqnarray}
&&  \delta \rho^{(j)} (\vec{q},\omega)=   
  -q\left(1-\frac{\omega^2}{q^2}\right) S_j \left[J_m^2 (0) D_\parallel^{(j)}  \right]  
                                  \delta {\cal V}^{(j)}_0    \nonumber \\
&&  + q S_j \left[ \frac{J_m^2 (0)}{E^{(0)}_j}  
  D_\parallel^{(j)} \right]  g_s {M_j^\star}^{(0)} ~ \delta \phi  
\label{displongi}  \ .
\end{eqnarray} }}
In a completely analogous way, one obtains for the scalar density:
{\small{
\begin{eqnarray}
&&  \delta \tilde{\rho}^{(j)}_s (q,\omega)=   
  -q\left(1-\frac{\omega^2}{q^2}\right) S_j \left[\frac{J_m^2 (0)}{E^{(0)}_j} D_\parallel^{(j)}  \right]  
                               {M_j^\star}^{(0)}   \delta {\cal V}^{(j)}_0    \nonumber \\
&&  + q S_j \left[ \frac{J_m^2 (0)}{{E^{(0)}_j}^2}  
  D_\parallel^{(j)} \right]  g_s {{M_j^\star}^{(0)}}^2 ~ \delta \phi  
\label{displongisca}  \ .
\end{eqnarray} }}
From eqs. (\ref{Lind},\ref{displongi},\ref{displongisca}) and the equations relating the 
perturbations of the mesonic fields with the corresponding perturbations of the currents 
given in the 
appendix, eqs. (\ref{eommes}), one obtains the dispersion relations. We have a set of 
five equations involving the
number densities, $\delta \rho^{(p)}$, $\delta \rho^{(n)}$, $\delta \rho^{(e)}$, and the 
scalar densities $\delta \tilde{\rho}^{(p)}_s$, $\delta \tilde{\rho}^{(n)}_s$, resulting in
the following matrix:
{\small{
\begin{equation}
\left(\begin{array}{ccccc}
a_{11}&a_{12}&a_{13}&a_{14}&a_{15}\\
a_{21}&a_{22}& 0 &a_{24}&a_{25}\\
a_{31}& 0 &a_{33}& 0 &0\\
a_{41}&a_{42}&a_{43}&a_{44}&a_{45}\\
a_{51} &a_{52}&0& a_{54} &a_{55}\\
\end{array}\right)
\left(\begin{array}{c}
  \delta \rho^{(p)} \\    \delta \rho^{(n)} \\
     \delta \rho^{(e)} \\ \delta \tilde{\rho}^{(p)}_s \\ \delta \tilde{\rho}^{(n)}_s
\end{array}\right)
=0 \ .
\label{det}
\end{equation} }}
The eigenmodes $\omega$ of the system are the solutions which correspond to the roots of the 
determinant of the latter matrix.
In the appendix, a more detailed discussion and the explicit expressions for the elements of 
the dispersion matrix are given.
\subsubsection{Transverse mode}
The transverse mode corresponds to perturbations on the plane perpendicular to the direction
of the external magnetic field. The dispersion relations are obtained by taking
$q_\perp$=q, $q_\parallel=0$ and 
$\delta {\cal V}_y$ = $\delta {\cal V}_z$ =0 in eq. (\ref{disdensity}).
\begin{figure*}[htb!]
	\includegraphics[width=1.\linewidth]{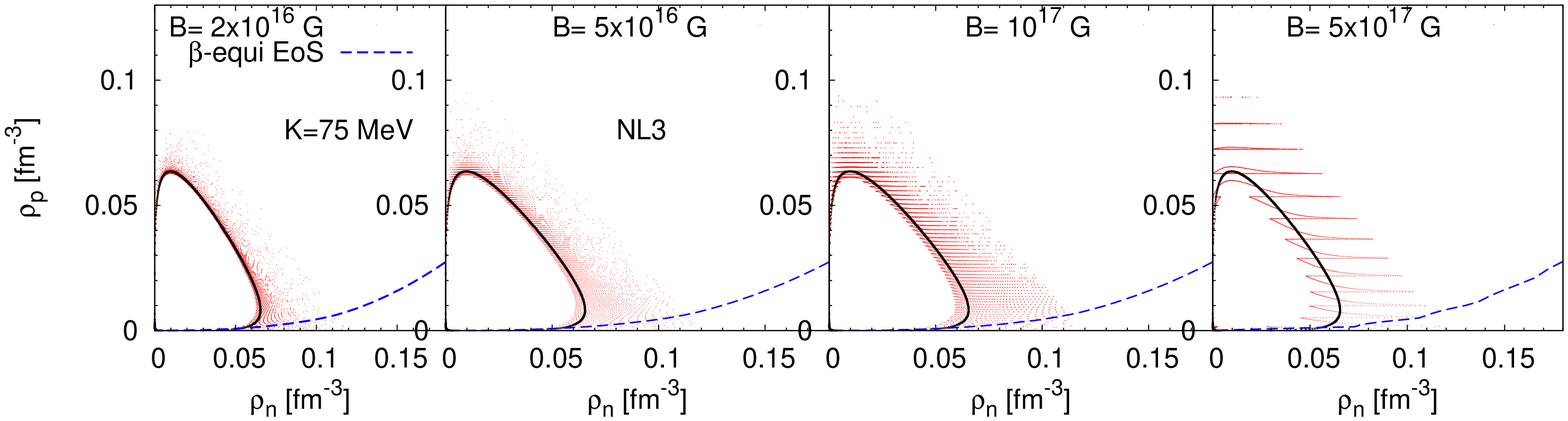}
	\includegraphics[width=1.\linewidth]{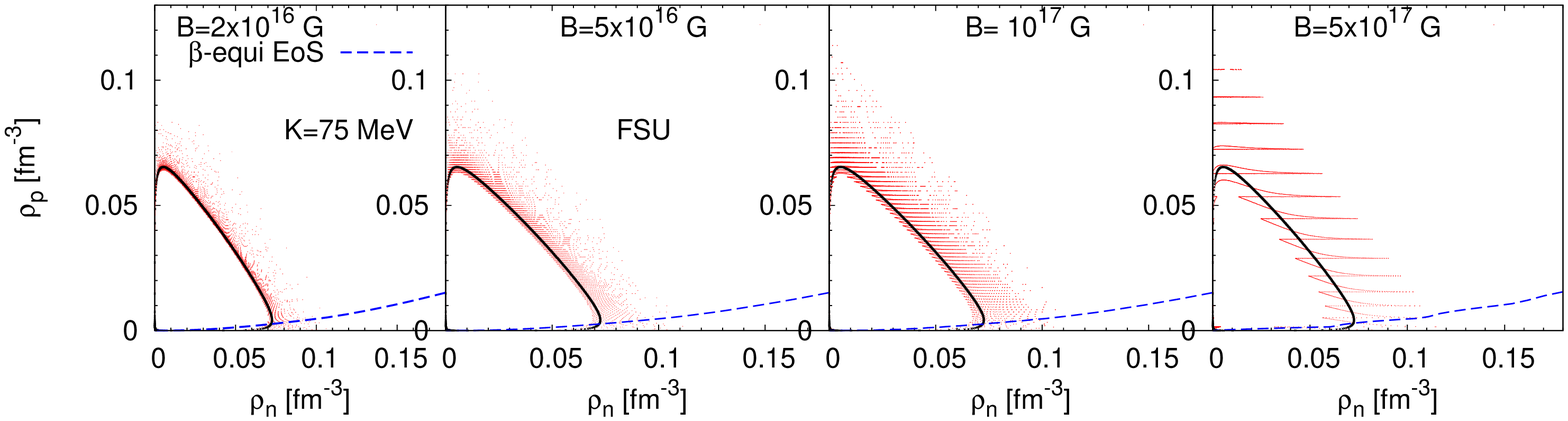}
	\caption{Dynamical spinodals for the NL3 (top panels) and the FSU (bottom panels)  parametrization, for several values of the magnetic field and for a moment transfer $k=75$MeV. The black curve corresponds to the $B= 0$ results.}
	\label{fig1:NL3}
\end{figure*}
Proceeding as in the longitudinal mode, one obtains:
{\small{
\begin{eqnarray}
\delta \rho^{(j)} (q, \omega) && = -q \left( 1 - \frac{\omega^{2}}{q^{2}} \right) S_{j}
\left[J_{m}^{2} (b) D_{\perp}^{(j)} \frac{m}{b}\right] \delta \mathcal{V}^{(j)}_{0} \nonumber \\
&& + q S_{j} \left[\frac{J_{m}^{2} (b)}{E_j^{(0)}} \frac{m}{b} D_{\perp}^{(j)}\right] g_{s} 
{M^\star}^{(0)} \delta \phi.
\label{eq1}
\end{eqnarray} }}
and
{\small{
\begin{eqnarray}
\delta \tilde{\rho}^{(j)}_s (q,\omega) = && -q \left( 1 - \frac{\omega^{2}}{q^{2}} \right) S_{j}
\left[\frac{J_{m}^{2} (b)}{E_j^{(0)}} D_{\perp}^{(j)} \frac{m}{b}\right] {M_j^\star}^{(0)} 
\delta \mathcal{V}^{(j)}_{0} \nonumber \\
&& + q S_{j} \left[\frac{J_{m}^{2} (b)}{{E_j^{(0)}}^{2}} \frac{m}{b}
D_{\perp}^{(j)}\right] g_{s} {{M_j^\star}^{(0)}}^{2} \delta \phi.
\label{eq2}
\end{eqnarray} }}
The dispersion relations for the transverse mode have the same matrix structure of eq. (\ref{det}), 
and the appropriate matrix elements for this case are discussed in detail in the Appendix.

\section{Results}

In the present section, we  discuss the effects of strong magnetic
fields on the structure of the inner crust of magnetars by analyzing
the dynamical spinodals for NL3 and FSU parametrizations. We consider
the propagation of both longitudinal and transverse modes, and define
the spinodal as the locus of the zero frequency solutions of the
dispersion relation obtained for each type. The two parametrizations
NL3 and FSU will be considered
because they represent two very distinct types of models: they have, respectively, a hard and a soft symmetry energy. We will also
calculate the maximum growth rates in the interior of the spinodal
surface and {the correspondents} wavelengths.  All  calculations are carried
out at zero temperature, and without taking in account the anomalous
magnetic moments (AMM) of nucleons. We will mostly consider magnetic field
intensities between $B=2\times10^{16}$G
and $B=5\times10^{17}$G, although we will also show some results for
$B\sim5\times 10^{18}$G.  The most intense fields detected on
the surface of a magnetar are not larger than $2\times10^{15}$G,
however, toroidal fields more intense than $10^{17}$G in stable
configurations were obtained in  Refs. \cite{kiuchi08, rezzolla12},
and, therefore, one may expect stronger fields in the interior of the
stars.

\subsection{Longitudinal modes and associated spinodal}

\begin{figure*}[htb]
	\includegraphics[width=1.\linewidth]{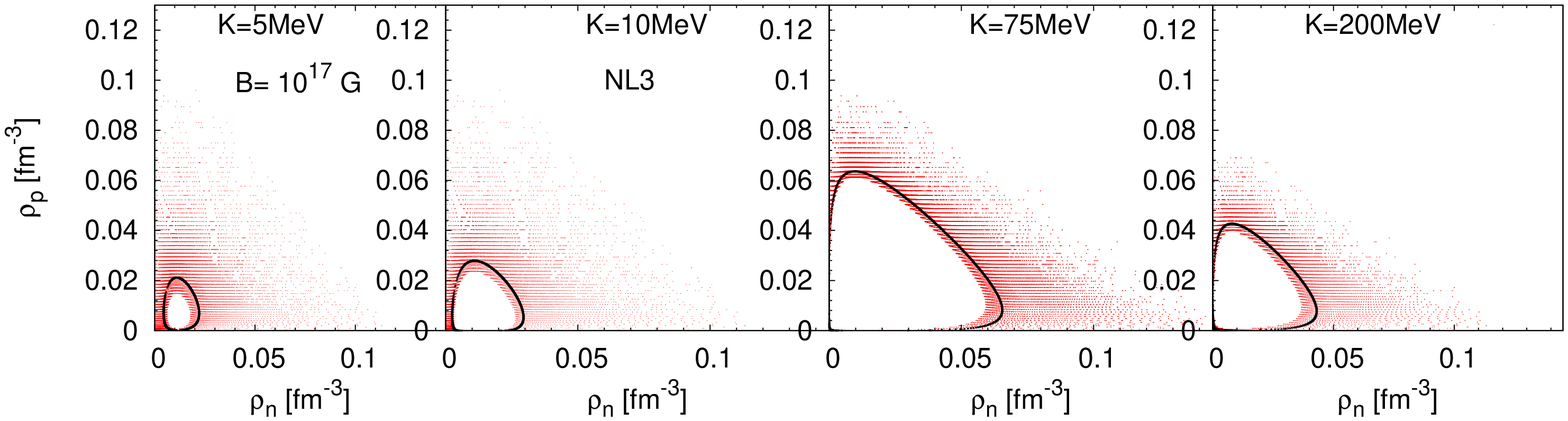}
        \includegraphics[width=1.\linewidth]{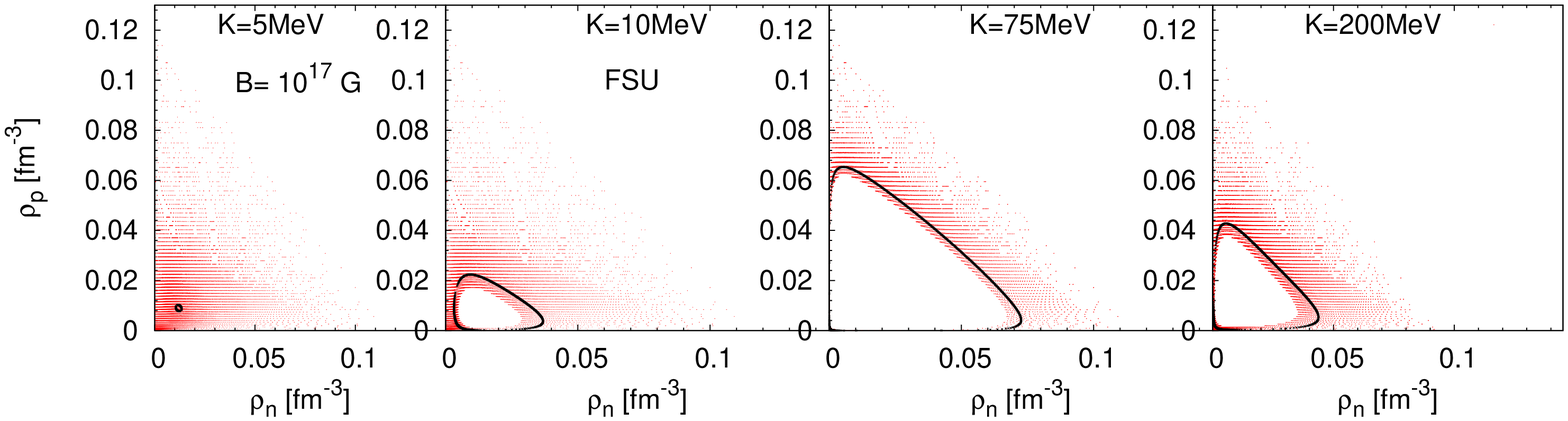}
	\caption{Dynamical spinodals for the NL3 (top panels) and the FSU (bottom panels) 
	parametrization,  for different moment transfer $k$ and for magnetic field $B=10^{17}$G. The black curve corresponds to the $B= 0$ results.}
	\label{fig2:NL3}
\end{figure*}

In Fig.\ref{fig1:NL3}, 
we show the dynamical
spinodal sections in the $(\rho_n, \rho_p)$ space for  magnetic field
intensities  $B=2\times10^{16}$G, $B=5\times10^{16}$G, $B=10^{17}$G
and $B=5\times10^{17}$G, obtained with the NL3 (top
panels) and  FSU (bottom panels) parametrizations.
The black lines represent the spinodal section at zero magnetic
field. The calculations were performed with $k = 75$ MeV, which is a
value of the transferred momentum that gives a spinodal section very
close to the envelope of the spinodal sections. Essentially, the size of
the spinodal section does not change much for $70<k<100$ MeV. In the
figure, we also
include the EoS of $\beta$-equilibrium matter represented by the blue
dashed line. This will allow us to estimate the crust-core transition
density  from the crossing of the
$\beta$-equilibrium EoS of stellar matter with the dynamical spinodal.

The spinodal sections have been obtained numerically by solving the
dispersion relation for $\omega = 0$. This was performed by looking
for the solutions at a fixed proton fraction and, for each solution, a
point was obtained. The solutions form a large connected region for
the lower proton and neutron densities, plus extra disconnected
domains that do not occur at $B = 0$. The spinodal sections appear
made of points, which, however, define close regions. The point-like
appearance of the sections is due to a numerical limitation. A higher
resolution in $(\rho_n, \rho_p)$ would complete the gaps. Similar
results have been obtained in \cite{Fang16}.

\begin{figure*}[htb!]
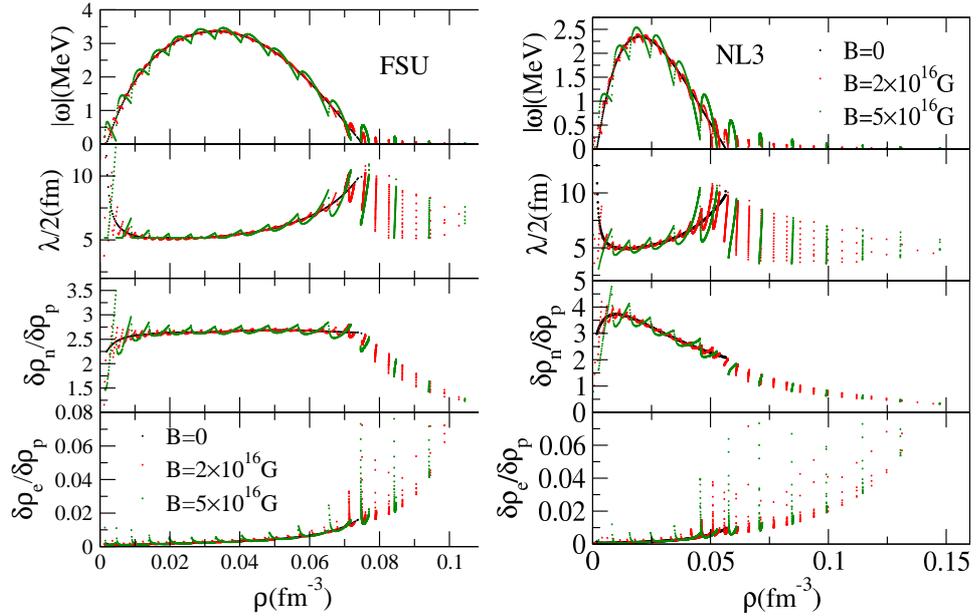

\begin{tabular}{cc}
	\includegraphics[width=0.35\linewidth]{fig3a.eps}
\includegraphics[width=0.35\linewidth]{fig3b.eps}
\end{tabular}
\caption{Largest growth rate $\Gamma=|\omega|$ (top panels), the
  corresponding half-wavelength (middle panels), the neutron-proton
  density fluctuation ratio versus density, and the electron-proton
  density fluctuation ratio (bottom panels) versus density for
  different magnetic-field intensities and matter with $y_p= 0.035$
  for FSU (left panels), and  with $y_p= 0.02$
  for NL3 (right panels). The black curve corresponds to the $B= 0$ results.}
	\label{fig3:NL3}
\end{figure*}

The structure of the spinodal section obtained for the strongest field
considered, $B =5\times 10^{17}$, clearly shows the effect of the
Landau quantization, as already shown in Ref.~\cite{Fang16}: there
are instability regions that extend to much larger densities than the
$B = 0$ spinodal section, while there are also stable regions that at
$B = 0$ would be unstable. This is due to the fact that the energy
density becomes softer, just after the opening of a new Landau level,
and harder when the Landau level  is most filled. For  weaker fields, the same
structure is found, but at a much smaller scale, due to the
increase of the number of Landau levels; see for $B =10^{17}$,
$B =5\times 10^{16}$ and $B =2\times 10^{16}$. The spinodal section
tends to the $B = 0$ one, as the magnetic field intensity is
reduced. The results obtained with the present approach coincide with
the ones obtained without the AMM in \cite{Fang16}. The present
approach, however, will allow us to define the propagation of modes in
an arbitrary direction with respect to the magnetic field
direction. In the next section, we will discuss transverse modes.

 Fig.~\ref{fig1:NL3} allows a comparison of the spinodal obtained for
 two models with a very different behavior of the symmetry energy: the
 symmetry energy slope at saturation is 118 MeV for NL3, and 60 MeV for
 FSU.  As discussed in previous works \cite{Fang16,Fang16a,chen17}, in
 the low proton density, the NL3 spinodal extends
 to larger neutron densities than FSU. On the other hand, at low
 neutron densities, the contrary occurs for FSU. These behaviors
 reflect the symmetry energy properties of both models: the larger
 NL3 slope at saturation density implies a smaller symmetry energy
 than FSU at   densities below $\rho\sim 0.1$ fm$^{-3}$, and this energetically favors asymmetric
  matter in NL3 with respect to FSU in that range of densities. Above
  $\rho\sim 0.1$ fm$^{-3}$, it is FSU that favors more asymmetric
  matter, and this explains the behavior of the spinodals for low
  neutron densities, when many Landau levels are occupied by protons.
For low proton densities, only the first Landau level is  being
occupied, and  the spinodal extends to larger neutron densities for NL3
with a large symmetry energy slope \cite{Rabhi2009,chen17,Fang16}.

\begin{figure*}[htb]
\begin{center}
\begin{tabular}{c}
\includegraphics[width=0.99\linewidth]{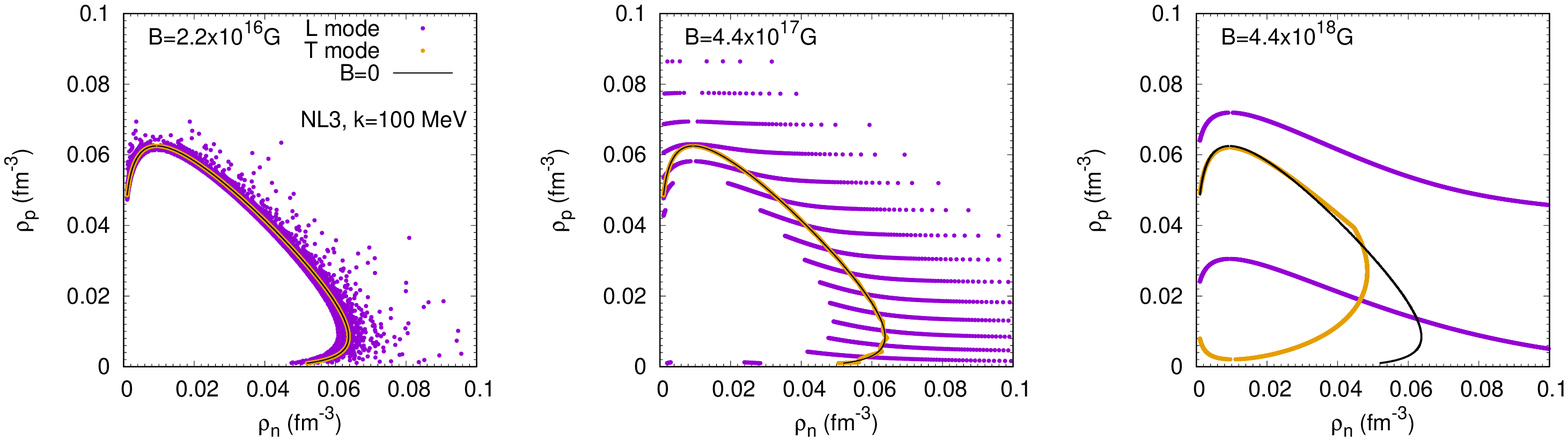}
\end{tabular}
\caption{Dynamic spinodals for {\it npe} matter subjected to an external magnetic field. The spinodal sections for both  longitudinal  (blue) and transverse modes (green) are given  for  $k = 100$ MeV and  a) $B = 2.2 \times 10^{16}$ G, b) $B = 4.4\times 10^{17}$ G, c)$B = 4.4 \times 10^{18}$ G. The black curve corresponds to the $B= 0$ results.}
\label{fig4}
\end{center}
\end{figure*}

In Fig. \ref{fig2:NL3}, we analyse the effect of the wave number on the
extension of the spinodal. As discussed in \cite{umodes06,umodes06a},
the extension of the spinodal increases when the wave number increases
from zero to a value of the order of $\sim 70-100$ MeV, due to the
effect of the Coulomb field. { We  see that, for all the values of the transferred momentum, $k$, the disconnected regions are present, and have a similar extension, i.e the upper limits of the disconnected region seem to be $k$-independent. This can be understood if one associates the isolated regions of instability to the onset  of a new Landau level which gives a particular extra stability as discussed in \cite{Rabhi2009}.}

For very long wavelengths, the electrons
and protons have to stick together in order to minimize the free
energy, and this reduces a lot the unstable region. If $B=0$, and for a model with a
large slope $L$, an unstable region at $k\sim 0$ still survives in the presence of electrons,
however, this is not anymore true for a model with a soft symmetry
energy like FSU, for which there is no unstable region. At finite $B$,
the effect of the Landau levels is important, and regions of
instability arise in a large region of the phase space, even for small
wave numbers. However, taking into account the discussion of Ref. \cite{Fang17}, it is expected that most of these effects are washed out for temperatures of the order of $\sim 0.5-1$ MeV. At large
wave numbers, $k>100$ MeV, the spinodal region reduces due to the
finite range of the nuclear force, as discussed in \cite{umodes06}.

 We will complete our discussion on the instabilities that arise due
 to the excitation of longitudinal modes, by calculating the largest
 growth rates for each density in  the unstable region. The
 corresponding modes are the ones that most probably drive the system
 into a non-homogeneous configuration. We consider that half the wavelength
 associated to these modes defines the average size of the clusters
 that are formed \cite{avancini08,pais10}, as discussed in \cite{Fang16,Fang16a}.

In Fig.~\ref{fig3:NL3}, 
the largest growth rates (top panels), the corresponding half
wavelength (second panels), the ratio $\delta \rho_n/\delta \rho_p$
between the neutron and proton density fluctuations (third panels),
and the ratio $\delta \rho_e/\delta \rho_p$ between the electron and
proton density fluctuations (bottom panels) are shown for two values
of magnetic fields $B = 2\times 10^{16}G$ (red) and $5\times10^{16}$G
(green), and for the NL3 and FSU models. For each one, the calculation is
done respectively at $y_p=0.02$ (NL3), and $y_p=0.035$ (FSU), which are
the average proton fractions obtained in a Thomas-Fermi calculation of
the pasta phases close to the crust-core transition \cite{grill12}.
 In all figures, the $B = 0$ results are plotted by a black curve. The results presented are
equivalent to the ones given in Fig. 4 of Ref. \cite{Fang16a} for the NL3$\omega\rho$
model, excluding the AMM contribution. As discussed there, the plots
present two well defined density regions: a low density region where
the different quantities fluctuate around the $B=0$ results, the
deviations being larger for the larger fields, and a region at higher
density, with no instabilities at $B=0$, but that, at $B>0$, present
regions of instability intercalated with stable regions. As expected
from the spinodal regions plotted in Fig. \ref{fig1:NL3}, the second
region extends to larger neutron densities for FSU, while the first
region is smaller for the NL3 model. Otherwise, as discussed in
\cite{Fang16a}, in the second region,  the clusters become
smaller, and more proton rich, as the density increases. In the bottom panel of
Fig.~\ref{fig3:NL3}, the ratio  $\delta \rho_e/\delta
\rho_p$ indicates how independent are the proton and electron
fluctuations. For all the densities shown, it is clear that electron
fluctuations are much smaller than the proton ones, and the system is
far from a scenario where electrons have to fluctuate in phase with
protons to lower the Coulomb energy, and which occurs for small
wave numbers.

\subsection{Transverse modes}
In the previous section, only longitudinal modes have been
considered. In the presence of strong magnetic fields, the motion of
charged particles is favored along the magnetic field direction, and,
therefore, it is expected that this is reflected on the spinodal
region obtained from  transverse modes.
In Fig. \ref{fig4},  we plot the solutions of the dispersion relations
for $\omega=0$, for both longitudinal and transverse modes.  We
consider three  different strengths of the external magnetic field: $B = 2.2 \times 10^{16}$ G, $B = 4.4\times 10^{17}$ G, and $B = 4.4 \times 10^{18}$ G. The transferred momentum $k$ has been fixed at 100 MeV. The present
comparison is undertaken only for the NL3 model (only values with
$\rho_n\le 0.1$ fm$^{-3}$ are shown). For FSU we obtain
similar results.
Some conclusions are in order: for the most intense field considered,
the spinodal connected to the longitudinal modes has two well defined
regions connected to the filling of the two Landau levels. As seen before, the
spinodal extends to quite large neutron densities. A completely {different} effect
is observed for the spinodal connected to the transverse modes. In this
case, the very strong field $B = 4.4 \times 10^{18}$ G gives rise to a
reduction of the spinodal at large neutron densities and isospin
asymmetries. The magnetic field hinders the propagation of
perturbations in a direction perpendicular to the magnetic
field. However, for a magnetic field one order of magnitude smaller, this
effect is washed out, and the transverse mode
spinodal almost coincides  with the $B=0$ spinodal. For this field
intensity, the longitudinal mode spinodal is still far from the $B=0$ limit.

\section{Conclusions}

In the present work, we have applied the covariant formulation of the
Vlasov equation presented in \cite{Heinz} to hadronic matter
described within a relativistic mean field approach. Within this
formalism, we have calculated the normal modes of  a
relativistic plasma, having in mind the study of magnetized nuclear
matter as found inside magnetars. The dispersion relations 
for both longitudinal and transverse density fluctuations were
determined for nuclear matter described within a relativistic
mean-field nuclear model with constant coupling constants.

In order to study the influence of the magnetic field on the extension
of the crust of a magnetized star, we have calculated the unstable
modes for  two different nuclear
models, which were chosen for their different density dependence  of
the symmetry energy.  The spinodal section, the locus of the zero
frequency modes, was determined for the longitudinal and transverse
density modes. We have also studied the growth-rates of the
unstable modes inside the spinodal sections, obtained by the largest
growth rates, e.g. the modes that most probably determine the evolution of
the system into a non-homogeneous phase, and discussed the effect of
the magnetic field on the size of the crust and on the size and
proton content of the clusters formed in the inner crust of the star.

We have recovered previous results obtained for the longitudinal mode
spinodal and corresponding unstable modes obtained with a generating function method.  In particular, the effect of the magnetic field on the
crust-core transition inside a neutron star was discussed. As already discussed
before, it was shown that this transition occurs at larger densities,
more strikingly if the slope of the symmetry energy is large, and that
the transition region has a finite density width and is characterized
by alternating stable and unstable regions.

However, within the formalism  developed in the present study,  we were able to study
the propagation of modes not necessarily aligned with the magnetic
field, and this was not possible in the approach of \cite{Fang16}. As an example, we have discussed the spinodal region that
originates from the excitation of transverse modes. For magnetic fields
of the order of $B\sim 5\times 10 ^{18}$ G, it was shown that the
spinodal section is much smaller than the $B=0$ section in the low
proton density range. We found, however, that this
effect is already washed out for a field one order of magnitude
smaller, and that for $B<5\times 10 ^{17}$ G,  the transverse mode
spinodal essentially coincides with the $B=0$ section.

Within the formalism developed, we aim at studying transport
coefficients, e.g. heat and electrical conductivities and viscosity
coefficients, of magnetized nuclear matter inside neutron stars. This
are important quantities to determine the magnetic field evolution
\cite{lander2016}, the  cooling of the star, or to
understand the gravitational driven r-mode instabilities \cite{haskel2015,kantor2017}.
 Another problem of interest for neutron stars is the  study of the effect of the magnetic
field on the non-homogeneous matter that constitutes the inner crust,
in particular, its effect on the shape of the clusters and orientation
of the pasta phases. {Although the dynamical spinodal includes the effect of surface tension (finite range of the force) and the Coulomb field, it does not take into account adequately the geometry, which is essential in this frustration problem of interplay between the Coulomb force and the surface tension. However, in the extended crust, we are dealing with quite dense matter, and it was seen that the wavelengths associated with the most unstable mode decrease outside the $B=0$ spinodal region. This may indicate that there will form thin tubes or deformed bubbles.} In other systems that present pasta like phases, such as
the surfactant liquid crystals, it has been shown that a magnetic
field may induce orientational phase transitions \cite{surfactant}.

\section*{ACKNOWLEDGMENTS}

This work was supported by Funda\c c\~ao para a Ci\^encia e Tecnologia, Portugal, 
under the projects UID/FIS/04564/2016 and POCI-01-0145-FEDER-029912 with  financial support  
from  POCI,  in  its FEDER  component,  and  by  the FCT/MCTES  budget through  national  funds  (OE).
Partial support comes from PHAROS COST Action CA16214, and from
Conselho Nacional de Desenvolvimento 
Cient\'{\i}fico e Tecnol\'{o}gico (CNPq) grant 6484/2016-1, and as a part of the
project INCT-FNA (Instituto Nacional de Ci\^encia e
Tecnologia - F{\'\i}sica Nuclear e Aplica\c c\~oes (INCT-FNA))
464898/2014-5 (SSA), and Coordena\c c\~ao de Aperfei\c coamento de Pessoal de Nivel Superior (CAPES) (BPB). H.P. is supported by FCT (Portugal) under Project No. SFRH/BPD/95566/2013.

\section*{Appendix}

\subsection{Equations of motion for the meson and eletromagnetic fields}
The equations of motion for the mesons and electromagnetic fields follow
from using the Euler-Lagrangian equations (\ref{mesonlag}):
{\footnotesize{
\begin{eqnarray}
 && \partial_t^2 \phi -\nabla^2 \phi + m_s^2 \phi +  \frac{\kappa}{2} \phi^2 +
 \frac{\lambda}{6} \phi^3 = g_s \sum_{j=p,n} \rho_s^{(j)}  \nonumber \\
 &&  \partial_t^2 V_{\mu} -\nabla^2 V_{\mu} + m_v^2 V_{\mu} + \frac{\xi}{6} V_\nu V^\nu V_\mu +
 2 \Lambda_v b_\nu b^\nu V_\mu
  = g_v \sum_{j=p,n} J^{(j)}_\mu  \nonumber \\
 &&  \partial_t^2 b_{\mu} -\nabla^2 b_{\mu} + m_\rho^2 b_{\mu} +
 2 \Lambda_v V_\nu V^\nu b_\mu
  = \frac{g_\rho}{2} \sum_{j=p,n} \tau_j J^{(j)}_\mu  \nonumber \\ 
  &&  \partial_t^2 A_{\mu} -\nabla^2 A_{\mu} 
  = e( J^{(p)}_\mu - J^{(e)}_\mu) =  \sum_{j=p,e} Q_j J^{(j)}_\mu  \nonumber  \ . 
\end{eqnarray}  }}
Now, we consider small deviations from the equilibrium in the fields, as given in eq. (\ref{smalldev}),
and perform a Fourier transform, obtaining:
{\footnotesize{
\begin{eqnarray}
 && \left[ -\omega^2 + {\vec q}^{~2} + {\tilde m}_s^2 \right] \delta \phi (\vec{q},\omega)
= g_s \sum_{j=p,n} \frac{2 M_j^{\star(0)}}{(2\pi)^3}\int \frac{d^3 p}{E^{(0)}_j}\delta f^{(j)}
\ , \nonumber \\
 && \left[ -\omega^2 + {\vec q}^{~2} + m_v^2 + \frac{\xi}{6}{V^{(0)}_0}^2 +
             2 \Lambda_v {b^{(0)}_0}^2 \right] \delta V_{\mu} +\frac{\xi}{3} {V_0^{(0)}}^2 \delta V_0 ~ 
\delta_{\mu 0} \nonumber \\ 
&& + 4 \Lambda_v V_0^{(0)} b^{(0)}_0 \delta b_0 ~\delta_{\mu 0}
  = g_v  \sum_{j=p,n} \frac{2}{(2\pi)^3}\int \frac{d^3 p}{E^{(0)}_j} p^\mu \delta f^{(j)} 
                                                            \ ,     \nonumber \\
 && \left[ -\omega^2 + {\vec q}^{~2} + m_\rho^2 +
             2 \Lambda_v {V^{(0)}_0}^2 \right] \delta b_{\mu}  \nonumber \\ 
&& + 4 \Lambda_v V_0^{(0)} b^{(0)}_0 \delta V_0 ~\delta_{\mu 0}
  = \frac{g_\rho}{2}  \sum_{j=p,n} \tau_j \frac{2}{(2\pi)^3}
  \int \frac{d^3 p}{E^{(0)}_j} p^\mu \delta f^{(j)} 
                                                            \ ,     \nonumber \\                                                                                                                        
  && \left[ -\omega^2 + {\vec q}^{~2} \right] \delta A_{\mu} 
   =  \sum_{j=p,e} Q_j \frac{2}{(2\pi)^3}
  \int \frac{d^3 p}{E^{(0)}_j} p^\mu \delta f^{(j)}    \ ,  \label{eommes}
\end{eqnarray} }}
where $\tau_{j}=\pm 1$ is the isospin projection for protons and neutrons, respectively, and 
the effective scalar mass is given by:
{\small{
\begin{equation}
 {\tilde m}_s^2 = m_s^2 + \kappa \phi^{(0)} + \frac{\lambda}{2}{\phi^{(0)}}^2 - g_s^2
 \sum_{j=p,n} d \rho^{(0)(j)}_s \ .
\end{equation} }}
\subsection{Dispersion relation matrix elements for longitudinal modes}
For the calculation of the dispersion relations for the longitudinal modes, 
eqs. (\ref{displongi},\ref{displongisca}), 
we first obtain $S_j$, by taking the derivative of the equilibrium 
distribution function, eq. (\ref{wignereq5}),
{\small{
\begin{eqnarray}
 &&    D_\parallel^{(j)} = \frac{\partial}{\partial p_\parallel} f^{(0)(j)}  = 
  \sum_{n=0}^\infty  
            \left[ L_{n}(2w^2)-L_{n-1}(2w^2) \right]  \nonumber \\ 
 &&\times (-1)^{n}~ e^{-w^2} (-\delta(p_\parallel -p_F^{(j)}(n)) 
 +\delta(p_\parallel +p_F^{(j)}(n)) )
 \ ,   \label{dpar} \nonumber
\end{eqnarray} }}
with $p_F^{(j)}(n)=\sqrt{\tilde{\mu}_j-{M_j^{\star(0)}}^2-2eBn}$. From eq. (\ref{Lind}), the latter 
equation, and using the property of the Bessel 
function, $J_m(b=0)=\delta_{m 0}$, we find:
{\footnotesize{
\begin{eqnarray}
 && S_j \left[ \frac{J_m^2 (0)}{{E^{(0)}_j}^k}  
  D_\parallel^{(j)} \right] = \frac{1}{q} \frac{1}{2\pi^2} \sum_{n=0}^{n_{max}} eB
     \int_0^\infty du~  \frac{e^{-u}}{E_{jn}^{(0) k+1}}(-1)^{n+1}   \nonumber \\
     && \times (L_n(2u)-L_{n-1}(2u)) 
 \left( \frac{p_F^{(j)}(n)}{ \left( \frac{\omega}{q}\right)^2 -
 \left(\frac{p_F^{(j)}(n)}{E_{jn}^{(0)}}\right)^2 } \right) \ ,\label{sj}
\end{eqnarray} }}
where $E_{jn}^{(0)}=\sqrt{M^{\star (0)}_j+{p_F^{(j)}}^2(n)+eBu}$.
The explicit matrix elements of eq. (\ref{det})) are:
{\small{
\begin{eqnarray*}
&&a_{11} =1+ F^{pp} {\rm L}^{(0)}_p ~,~ a_{12}= F^{pn} {\rm L}^{(0)}_p ~,~
a_{13} = B_A {\rm L}^{(0)}_p  ~,~    \\
&&a_{14}  =  M^{\star (0)} B_s {\rm L}^{(1)}_p ~, ~
a_{15} =  M^{\star (0)} B_s {\rm L}^{(1)}_p \\
&& a_{21}=F^{np} {\rm L}^{(0)}_n ~,~a_{22}=1+ F^{nn} {\rm L}^{(0)}_n ~,~ a_{23}=0 ~,~ \\
&& a_{24} = M^{\star (0)} B_s {\rm L}^{(1)}_n ~,~ a_{25}= M^{\star (0)} B_s {\rm L}^{(1)}_n \\
&& a_{31} = B_A {\rm L}^{(0)}_e ~,~ a_{32}= 0 ~,~ a_{33}= 1- B_A {\rm L}^{(0)}_e ~,~ a_{34}=0 ~, \\
&&a_{35}=0 \\
&& a_{41} = F^{pp} M^{\star (0)}  {\rm L}^{(1)}_p ~,~ a_{42}= F^{pn} M^{\star (0)}  {\rm L}^{(1)}_p ~,~ 
 \\
&& a_{43}= B_A M^{\star (0)} {\rm L}^{(1)}_p ~,~ 
a_{44}=1+ B_s F^{pp} {M^{\star (0)}}^2 {\rm L}^{(2)}_p ~,~ \\
&& a_{45}=B_s F^{pp} {M^{\star (0)}}^2 {\rm L}^{(2)}_p \\
&&  a_{51} = F^{np} M^{\star (0)}  {\rm L}^{(1)}_p ~,~ a_{52}= F^{nn} M^{\star (0)}  {\rm L}^{(1)}_n 
~,~a_{53}= 0 ~,~ \\
&& a_{54}=B_s {M^{\star (0)}}^2 {\rm L}^{(2)}_n ~,~ a_{55}= 1+ B_s {M^{\star (0)}}^2 {\rm L}^{(2)}_n \ ,
\end{eqnarray*} }}
where 
\begin{equation}
 F^{ij}=-\left( B_v+\tau_i \tau_j B_\rho + \frac{Q_i Q_j}{e^2} B_A \right) \nonumber
\end{equation}
and
{\small{
\begin{eqnarray}
&& B_v = \frac{1}{2\pi^2}\frac{g_v^2}{\omega^2-\omega_v^2} \left( 1 - \frac{\omega^2}{q^2} \right) \nonumber \\
&& B_\rho = \frac{1}{2\pi^2}\frac{(g_\rho/2)^2}{\omega^2-\omega_\rho^2} 
\left( 1 - \frac{\omega^2}{q^2} \right) \nonumber \\
&& B_s = \frac{1}{2\pi^2}\frac{g_s^2}{\omega^2-\omega_s^2} ~,~B_A = -\frac{e^2}{2\pi^2}\frac{1}{q^2} \nonumber \ ,
\end{eqnarray} }}
with $\omega_s^2=\tilde{m}_s^2+ \vec{q}^{~2}$, $\omega_v^2=m_v^2+ \vec{q}^{~2}$, 
and $\omega_\rho^2=m_\rho^2+ \vec{q}^{~2}$. The generalized Lindhard functions for the proton and 
electron are defined as:
{\small{
\begin{eqnarray}
 && {\rm L}_j^{(k)} =  eB \sum_{n=0}^{n_{max}}
     \int_0^\infty du(L_n(2u)-L_{n-1}(2u))  \nonumber \\
     && \times \frac{e^{-u}}{{{E_{jn}^{(0) k+1}}}}(-1)^{n+1} 
 \left( \frac{p_F^{(j)}(n)}{ \left( \frac{\omega}{q}\right)^2 -
 \left(\frac{p_F^{(j)}(n)}{E^{(0)}_{jn}}    \right)^2 } \right) \ . \nonumber
\end{eqnarray} }}
Notice that we have defined the generalized Lindhard function, ${\rm L}_j^{(k)}$, such 
that the present Vlasov magnetized dispersion relation and the corresponding non-magnetized 
Vlasov dispersion relation of 
Ref. \cite{avancini05} differs only in what concerns this function. Hence, the limit $B=0$ is
easily obtained.
\subsection{Dispersion relation matrix elements for transverse modes}
An equation similar to eq. (\ref{sj}) provides the necessary solutions for the various $S_{j}$ 
functions. However,
an important complication arises in the context of the transverse modes, in that the
argument of the Bessel functions
is not zero. This means one needs to compute the contribution of the Bessel functions
and perform the integrals:
{\small{
\begin{eqnarray}
 && S_j \left[ \frac{J_m^2 (b)}{{E^{(0)}_j}^k} \frac{m}{b} 
  D_\perp^{(j)} \right] = \frac{1}{2\pi^2} \sum_{m=-\infty}^{\infty} 
  \int_{-\infty}^{\infty} d p_\parallel \nonumber  \\
     && \times \int_{0}^{\infty} dp_{\perp} p_{\perp}  \frac{m}{b}
     \frac{J_{m}^{2} (b)}{{E_j^{\star(0)k}}}   \frac{  D_{\perp}^{(j)}}
     {\omega + \frac{Q_{j} B m}{{E_j^{\star(0)k}}}} ~ .
       \label{eq3}
\end{eqnarray} }}
Next, we obtain $S_j$ by taking the derivative of the equilibrium 
distribution function, eq. (\ref{wignereq5}),
{\small{
\begin{eqnarray}
 &&    D_\perp^{(j)} = \frac{\partial}{\partial p_\perp} f^{(0)(j)}  = 
  \sum_{n=0}^\infty  
            \left[ L_{n}(2w^2)+L_{n-1}(2w^2) \right]  \nonumber \\ 
 &&\times (-1)^{n+1}~ e^{-w^2}~ \frac{2 p_\perp}{eB}~ \theta (p_F^{(j)}(n)-p_\parallel) 
 \ ,  \label{dperp}
\end{eqnarray} }}
and substituting the latter result in eq. (\ref{eq3}). After some simple manipulations, 
by separating negative and positive (as well as zero) values of $m$,
and using properties of the Bessel functions in order to reduce the integrals to more usual forms,
one obtains:
{\footnotesize{
\begin{eqnarray}
 && S_j \left[ \frac{J_m^2 (b)}{{E^{(0)}_j}^k} \frac{m}{b} 
  D_\perp^{(j)} \right] = \frac{1}{2\pi^2} \frac{1}{q} \sum_{n=0}^{n_{max}} 
  \int_{0}^{p_F^{(j)}(n)} d p_\parallel \nonumber  \\
     && \times 8(eB)\int_{0}^{\infty} dp_\perp p_\perp 
      \left[ L_{n}(2w^2)+L_{n-1}(2w^2) \right](-1)^{n+1}~ e^{-w^2}~ \nonumber \\
     && \times  \frac{1}{{E_j^{\star(0)}}^{k+1}}  \sum_{m=0}^{\infty} (1-\frac{\delta_{m0}}{2})
     \frac{ m^2~J_m(b)^2}
     {\omega^2 - \frac{(Q_{j} B m)^2}{{E_j^{\star(0)}}^{2}}} 
       \equiv \frac{1}{q} \frac{1}{2\pi^2} L_j^{(k)} ~ .
       \label{eq10}
\end{eqnarray} }}
These equations simplify considerably in some particular cases of interest. For instance, to determine the spinodal boundary, we impose $\omega=0$.
Below, we show the results for $L_j^{(k)}$ in this particular case:

{\small{
\begin{eqnarray}
&&L^{(k)}_{j} = \frac{8}{eB} \int_{0}^{\infty} dp_{\perp} p_{\perp} 
\frac{(-1)^{n} e^{-w^{2}}}{{E_j^{\star(0)}}^{k-1}}
\sum_{n=0}^{n_{max}}  \int_{0}^{p_F^{(j)}(n)}  dp_\parallel  \nonumber   \\
&& \times (L_n(2u)+L_{n-1}(2u)) \sum_{m=0}^{\infty}
(1-\frac{\delta_{m0}}{2}) J_{m}^{2} (b) \nonumber  ~.
\end{eqnarray} }}
The latter equation can be simplified further by using the following
Bessel function property \cite{grad}:
{\small{
\begin{equation}
 \sum_{m=0}^{\infty}
(1-\frac{\delta_{m0}}{2}) J_{m}^{2} (b) = \frac{1}{2} ~.
\end{equation} }}
The special cases involved in the calculation of the dispersion relations are the following:
{\footnotesize{
\begin{eqnarray}
L^{(0)}_{j} =&&\frac{2}{eB}\sum_{n=0}^{n_{max}} \int_{0}^{\infty} dp_{\perp} p_{\perp} 
(L_n(2w^2)+L_{n-1}(2w^2)) (-1)^{n} e^{-w^{2}}\nonumber \\
&& \times \left[p_{F}^{(j)} (n) 
\sqrt{{M_{j}^{\star(0)}}^{2} + p_{\perp}^{2} + {p_{F}^{(j)} (n)}^{2}} + \right. \nonumber \\
&& \left. + ({M_{j}^{\star(0)}}^{2} +  p_{\perp}^{2}) 
\ln \frac{p_{F}^{(j)} (n) + \sqrt{{M_{j}^{\star(0)}}^{2} + p_{\perp}^{2} + {p_{F}^{(j)}(n)}^{2}}}
{\sqrt{{M_{j}^{\star(0)}}^{2} + p_{\perp}^{2}}}   \right] \nonumber
\label{eq11}
\end{eqnarray} }}
{\footnotesize{
\begin{equation}
L^{(1)}_{j} = \frac{4}{eB}  \sum_{n=0}^{n_{max}} \int_{0}^{\infty} dp_{\perp} p_{\perp} 
(L_n(2w^2)+ L_{n-1}(2w^2)) (-1)^{n} e^{-w^{2}} {p_{F}^{(j)} (n)} \nonumber
\label{eq6}
\end{equation} }}
{\footnotesize{
\begin{eqnarray}
L^{(2)}_{j} = && \frac{4}{eB}  \sum_{n=0}^{n_{max}} \int_{0}^{\infty} dp_{\perp} p_{\perp} 
(L_n(2w^2)+L_{n-1}(2w^2)) (-1)^{n} e^{-w^{2}} \nonumber \\
&& \times \ln \frac{{p_{F}^{(j)}(n)}+\sqrt{{M_{j}^{\star(0)}}^{2}+p_{\perp}^{2} +
{p_{F}^{(j)} (n)}^{2}}}{\sqrt{{M_{j}^{\star(0)}}^{2} + p_{\perp}^{2}}} \nonumber
\label{eq7}
\end{eqnarray} }}
We have chosen the notation $L^{k}_{j}$ to mirror the generalized Lindhard functions, given in
Appendix B, and, in fact,
substituting the equations above for those functions, one ends with a matrix for the 
transverse modes that is identical
to the one obtained for the longitudinal modes, (\ref{det}).

\end{document}